\documentclass[aps,pra,twocolumn,showpacs,superscriptaddress]{revtex4}

\usepackage{amsfonts,amsmath,amssymb}
\usepackage{color,graphicx}
\usepackage[pdfstartview=FitH,colorlinks=true,linkcolor=blue,citecolor=blue,urlcolor=blue]{hyperref}

\begin{document}

\title{Confinement and precession of vortex pairs in coherently coupled Bose-Einstein condensates}
\author{Marek Tylutki}
\email{marek.tylutki@ino.it}
\affiliation{INO-CNR BEC Center and Dipartimento di Fisica, Universit\`a di Trento, Via Sommarive 14, I-38123 Povo, Italy}
\author{Lev P. Pitaevskii}
\affiliation{INO-CNR BEC Center and Dipartimento di Fisica, Universit\`a di Trento, Via Sommarive 14, I-38123 Povo, Italy}
\affiliation{Kapitza Institute for Physical Problems RAS, Kosygina 2, 119334 Moscow, Russia}
\author{Alessio Recati}
\affiliation{INO-CNR BEC Center and Dipartimento di Fisica, Universit\`a di Trento, Via Sommarive 14, I-38123 Povo, Italy}
\affiliation{Physik Department, Technische Universit\"at M\"unchen, James-Franck-Stra{\ss }e 1, 85748 Garching, Germany}
\author{Sandro Stringari}
\affiliation{INO-CNR BEC Center and Dipartimento di Fisica, Universit\`a di Trento, Via Sommarive 14, I-38123 Povo, Italy}
\date{\today}

\begin{abstract}
The dynamic behavior of vortex pairs in two-component coherently ({\it Rabi}) coupled
Bose-Einstein condensates is investigated in the presence of harmonic
trapping. We discuss the role of the surface tension associated with the
domain wall connecting two vortices in condensates of atoms occupying  different spin states and its effect
on the precession of the vortex  pair. The results, based on the
numerical solution of the Gross-Pitaevskii equations, are compared with the
predictions of an analytical macroscopic model and are discussed as
a function of the size of the  pair, the Rabi coupling and the
inter-component interaction. We show that the increase of the Rabi coupling results in the 
disintegration of the domain wall into smaller pieces, connecting vortices of new-created vortex pairs. The resulting scenario is the analogue of quark confinement and string breaking in quantum chromodynamics. 
\end{abstract}

\maketitle

\section{Introduction}
Vortices and solitons are among the most striking features of superfluids and superconductors. Recent experimental advances in cold atomic gases have made it possible to study multi-component systems with a vector order parameter \cite{UedaKurnRMP}. The nature of vortices and solitons in such systems is far from being a trivial extension of the single component case. In the context of Bose-Einstein condensates (BECs) the situation is very rich. Not only it is possible to have condensates with two or more components, but it is also feasible to tune the interaction strengths, which allows to explore the nature of vortices in different phases (see, for example, \cite{UedaReview,UedaKurnRMP,AftalionVortices}). Furthermore, if the components correspond to different spin states of the same atom, it is possible to produce coherent Rabi coupling between the components using \textit{rf} transitions, yielding novel topological features in the structure of vortices and solitons.

In this work we consider a mixture of two coherently coupled BECs. Some properties of these systems like coherent Rabi oscillations, internal self-trapping effects, and the ferromagnetic classical bifurcation have already been addressed experimentally~\cite{UntwistCornell1999,DipoleLargeRabiCornell2000,BifurcationZibold2010}. In the absence of Rabi coupling, a two-component Bose gas has a $U(1)\times U(1)$ gauge symmetry, which is broken in the condensed, superfluid phase. In this case vortices behave similarly to the single component case as long as the inter-component interaction is small enough and the mixture is miscible. In the presence of Rabi coupling, which produces coherent transitions between the spin components, the situation drastically changes. The gas has only a single $U(1)$ symmetry related to the conservation of the total particle number, and a gap opens in the spin channel, since the coupling locks the relative phase of the order parameter of different spin components. As  pointed out in the seminal paper by Son and Stephanov~\cite{Son2002} (see also~\cite{Garcia-Ripoll2002}), in coherently coupled uniform two-component superfluids, vortex lines in a single component cannot exist as single objects, but only in pairs of different spin components (the so called {\it 'half-vortices'}), which are  connected by a domain wall. The domain wall can be interpreted as a sine-Gordon soliton and is characterized by a $2\pi$  jump in the relative phase  of the two  spin components. Vortices can exist only in pairs because a single vortex would be attached to a domain wall of infinite area, with an infinite energy cost. Stationary configurations of vortex pairs (also known as {\it 'merons'}~\cite{UedaReview}) have already been investigated by solving the Gross-Pitaevskii (GP) equation in the presence of a rotating trapped two-component gas~\cite{Kasamatsu2004,Kasamatsu2005}. In the same spirit, numerical extensions to more than two coherently coupled BECs have been produced in recent years~\cite{Eto2012,Cipriani2013}. Pairs of half-vortices connected by a domain wall share many features with quark confinement in quantum chromodynamics (QCD). In particular, the surface energy related to the domain wall is proportional to its length and creates a distance-independent force that attracts vortices to each other. If the surface tension of the wall is large enough, it can break, creating new pairs of vortices similar to string breaking in QCD. 

In the present work, we provide an explicit investigation of the precession of vortex pairs emphasizing, in particular, the competitive role played by the the Magnus force acting on a moving vortex, and the attractive force due to the surface tension of the domain wall. Furthermore, we show explicitly that the domain wall connecting the vortex pair breaks for large values of the Rabi coupling. Our analysis is based on the 2D GP equations 
\begin{eqnarray}
i\hbar \partial _{t}\psi _{1} &=&(H_{0}+g|\psi _{1}|^{2}+g_{12}|\psi
_{2}|^{2})\psi _{1}-\frac{1}{2}\hbar \Omega _{\mathrm{R}}\psi_{2}~,\rule{0.7cm}{0cm} \notag \\
i\hbar \partial _{t}\psi _{2} &=&(H_{0}+g|\psi _{2}|^{2}+g_{12}|\psi
_{1}|^{2})\psi _{2}-\frac{1}{2}\hbar \Omega _{\mathrm{R}}\psi_{1}~,\rule{0.7cm}{0cm}
\label{GP}
\end{eqnarray}
for the order parameters $\psi _{1}$ and $\psi _{2}$ of the two components respectively. The two equations are coupled not only by the inter-component interaction term proportional to $g_{12}$, but also by the Rabi coupling $\Omega _{\mathrm{R}}>0$ between the two spin states. In the above equations $H_{0}=-\frac{\hbar ^{2}}{2m}\nabla ^{2}+\frac{1}{2}m\omega_{\perp }^{2}r_{\perp }^{2}$ is the single-particle Hamiltonian, and $r_\perp$ is the radial coordinate measuring the distance from the trap's center ($r_\perp^2 = {\bf r}^2$). 

The presence of the Rabi coupling implies that, at equilibrium, the  phases of the two components are equal: $\theta_1 = \theta_2$. In the following we will assume $g_{11} = g_{22} \equiv g$ and consider the miscible (paramagnetic) case, ensured by the inequality $g_{12} < g + \hbar \Omega_{R} / n$. Then the Rabi coupling implies that the equilibrium densities are equal: $n_1 = n_2 = n/2$, with $n_i = |\psi_i ({\bf r})|^2$, $(i = 1,2)$, and $n = n_1 + n_2$. 

\section{Thomas-Fermi model}
A simple macroscopic description of the dynamics of these novel
configurations is obtained by considering a vortex pair located symmetrically with respect to the center of the harmonic potential
and by writing the excess energy per particle with respect to the ground state
in the form 
\begin{equation}
E(d)=2E_{v}(d)+E_{wall}(d)+E_{int}(d) ~,
\label{Ed}
\end{equation}
where $d$ is the distance of each vortex from the center (and hence $2d$ is the distance separating the two vortices). In the above equation $E_{v}(d) = N E_{V}(1-d^{2}/R_\perp^{2})$ is the energy of a single vortex, for which one can use the macroscopic estimate $E_{V} \approx (\hbar^{2} / m R_\perp^{2}) [\ln(1.46 R_\perp / \xi) - \frac12]$ \cite{Lundh1997,BlackBook,Fetter2001}; $R_\perp$ is the Thomas-Fermi radius of the trapped atomic cloud, $N = \int d{\bf r}\, n({\bf r})$ is the number of atoms and $\xi =\hbar /\sqrt{2mgn}$ is the healing length, while 
\begin{subequations}
\begin{eqnarray}
E_{wall}(d) & = & \int_{-d}^{d}dr_{\perp }\,\sigma (r_{\perp },\Omega _{\mathrm{R}}) ~, \label{Ewall} \\
\sigma (r_{\perp },\Omega _{\mathrm{R}}) & = & 2^{3/2}\frac{\hbar ^{3/2}}{m^{1/2}}n(r_{\perp })\sqrt{\Omega _{\mathrm{R}}}~ \label{sigma}
\end{eqnarray}
\label{eq.tension}
\end{subequations}
is the surface energy associated with the domain wall connecting the two vortices. In the above equations $\sigma $ is the surface tension~\cite{Son2002} of the wall and $n(r_{\perp })$ is the density of the gas. In a uniform medium the energy $E_{wall}$ increases linearly with the distance $d$. Finally, $E_{int}(d)$ is the energy of the repulsive interaction between the two vortices fixed by the inter-species coupling constant $g_{12}$. This interaction energy was calculated analytically in \cite{Eto2011} in the absence of a coherent coupling ($\Omega_{\mathrm{R}}=0$).

Starting with the above equations, one can calculate the precession frequency
characterizing the rotation of the vortex pair as
\begin{equation}
\Omega _{\mathrm{prec}}=\frac{\partial E}{\partial L_{z}}=\frac{\partial E}{\partial d}\frac{\partial d}{\partial L_{z}} ~,
\label{precession}
\end{equation}
where the angular momentum $L_{z}$ of the vortical configuration is calculated in terms of the distance $d$ of each vortex from the center using the relation~\cite{BlackBook} $\langle L_{z}\rangle = N \hbar(1-d^{2}/R_\perp^{2})^{2}$, holding for 2D harmonically trapped condensates.

For a fixed value of the interaction parameters and the Rabi coupling $\Omega_{R}$, the precession angular velocity will depend on the distance $2d$. The above formalism is immediately generalized to the case of a trap rotating with angular velocity $\Omega_{\mathrm{rot}}$ by adding the term $-\Omega_{\mathrm{rot}} \langle L_z \rangle$ to the expression for the total energy.

In principle the same formalism would also allow for the determination of
the size of the pair at equilibrium (vortex molecule).
The corresponding condition is fixed by the equation $\partial E/\partial
d=0 $. Notice that this condition implies the absence of precession, as
immediately follows from (\ref{precession}). The occurrence of a stable
vortex molecule with $d\neq 0$ cannot be ensured in the absence of
the repulsive interaction term $E_{int}$ in Eq.(\ref{Ed}). In fact, if $E_{int}=0$, the
stationary solution $\partial E/\partial d=0$ always corresponds to a local
maximum of $E(d)$ rather than to a minimum. 

In order to make the role of Rabi coupling clearer, we first consider the case of the absence of inter-component interaction, i.e. we assume $g_{12}=0$. In this case the precession frequency in the absence of rotation ($\Omega _{\mathrm{rot}}=0$) results from the competition between $E_{wall}$ and $E_{v}$, which act in opposite directions. Assuming $d \ll R_\perp$ one finds 
\begin{equation}
\Omega _{\mathrm{prec}} = -\frac{\sigma }{\pi \hbar nd} + \frac{E_V}{\hbar} = -2\sqrt{\frac{2\hbar }{m}}\frac{\sqrt{\Omega_{\mathrm{R}}}}{\pi d} + \frac{E_V}{\hbar} 
\label{eq.prec}
\end{equation}
showing that, if $\Omega_{\rm R}$ (and hence $E_{wall}$) is sufficiently large the precession will proceed in the opposite direction with respect to the flow generated by the vortex line. In the limit $E_{wall}\gg E_{v}$, the precession frequency will be proportional to the surface tension of the wall and hence to $\sqrt{\Omega_{\mathrm{R}}}$. This result can also be obtained by considering the equilibrium between the Magnus force and the surface tension $\sigma$:  ${\mathbf \kappa} \times \mathbf{v}_{L}n/2+\sigma =0$, where ${\bf v}_{L}=\Omega _{\mathrm{prec}}d$ is the velocity of the vortex line with respect to the fluid. The vector $\mathbf \kappa$, with modulus $|{\mathbf \kappa}| = 2\pi \hbar / m$, is oriented  along the vortex line and  indicates the velocity circulation. In the calculation of the Magnus force we used the unperturbed value $n/2$ for the density of each component. 

The condition $E_{wall} \gg E_v$ is easily achieved experimentally. In the following we will consider a gas of sodium  atoms confined by  a harmonic potential with trap frequency $\omega_{\perp} / 2 \pi = 10\, \mathrm{Hz}$ and with a chemical potential $\mu = gn_{0}=50\,\hbar \omega_{\perp }$, where $n_{0}$ is  the central density. This corresponds to a Thomas-Fermi radius $R_{\perp} = 10\, \xi_{ho} = 66.5\,\mu m$, where $\xi_{ho} = \sqrt{{\hbar/m \omega_{\perp}}{}}$ is the harmonic oscillator length. In the presence of Rabi coupling $\Omega_{\mathrm{R}}/2\pi = 5\, \mathrm{Hz}$  one then finds that a vortex pair, separated by the distance  $2d=0.53\, R_{\perp }$, is characterized by the precession frequency $\Omega _{\mathrm{prec}}/2\pi \simeq -2\, \mathrm{Hz}$.  Setting $\Omega_{\mathrm{R}} = 0$ one instead would find the value $\Omega_{\mathrm{prec}}/2\pi \simeq +0.45\,\mathrm{Hz}$ of opposite sign.

The above model, including Eq.~(\ref{eq.tension}) for the surface tension, is justified as long as the width of the domain wall, fixed by the relationship $d_{wall}=2\sqrt{\hbar /m\Omega _{\mathrm{R}}}$, is much larger than the healing length $\xi = \hbar / \sqrt{2mgn} $, but much smaller than the size $2d$ of the vortex pair. If the width $d_{wall}$ of the domain wall is much larger than the size of the vortex pair (corresponding to very small values of $\Omega _{\mathrm{R}}$), the effect of the Rabi coupling should be treated using a perturbative approach~\cite{Fetternew}.

\section{Numerical results}
\begin{figure}
\centering
\includegraphics[width=.98\columnwidth, clip=true]{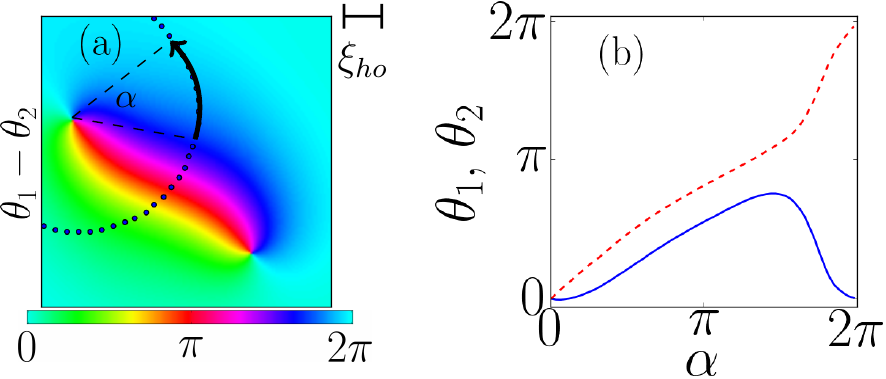}
\caption{(color online) (a) Relative phase $\protect\theta_1 - \protect\theta_2$ of the two components near a vortex pair in the presence of Rabi coupling $\Omega_{\mathrm{R}} = 0.5 \protect\omega_\perp$.  We can see that the phase jump between the two vortices is confined within the narrow domain wall stretching between the vortices. The dotted line shows the circle around which the phase is calculated in the right panel. (b) Phases $\theta_1$ (red dotted line) and $\theta_2$ (blue solid line) along a circle centered in the vortex of the first component. The phase $\theta_1$ makes a $2\protect\pi $ winding around a vortex, with half of the jump  concentrated in a short interval of the polar angle $\alpha$ (of the coordinate system centered in the vortex core, as shown in panel (a)). The phase $\theta_2$ is instead single valued.}
\label{fig12}
\end{figure}

In Fig.~\ref{fig12} we show the typical behavior of the relative phase of a
vortex pair in the presence of a 2D harmonic trap, obtained by
solving numerically the coupled GP Eqs.~(\ref{GP}) with  $g_{12}=0$. The structure of the domain wall is clearly visible. The result shown in the figure is obtained as follows: first, we
symmetrically imprint two vortices, one in each component, far away from the
trap's center; then we perform an imaginary time evolution, during which the
domain wall in the relative phase forms and the vortices start approaching
each other, with the energy of the system decreasing. We stop the
simulation at a certain point, in order to produce a vortex pair of the
desired size. This configuration serves as the initial condition for a
subsequent real time evolution, in which the pair exhibits  precession. The
configuration in Fig.~\ref{fig12} shows the phase after a short precession
time,  the small asymmetry in the shape of the wall being caused 
 by the rotation. 

Fig.~\ref{fig12}(b) shows explicitly the behavior of the phases of the two components calculated along the contour shown in Fig.~\ref{fig12}(a), i.e., around the vortex of the spin component 1. The figure clearly reveals the $2\pi$ jump in the relative phase $\theta_{1}-\theta_{2}$ near the domain wall. 

\begin{figure}
\centering
\includegraphics[width=.98\columnwidth, clip=true]{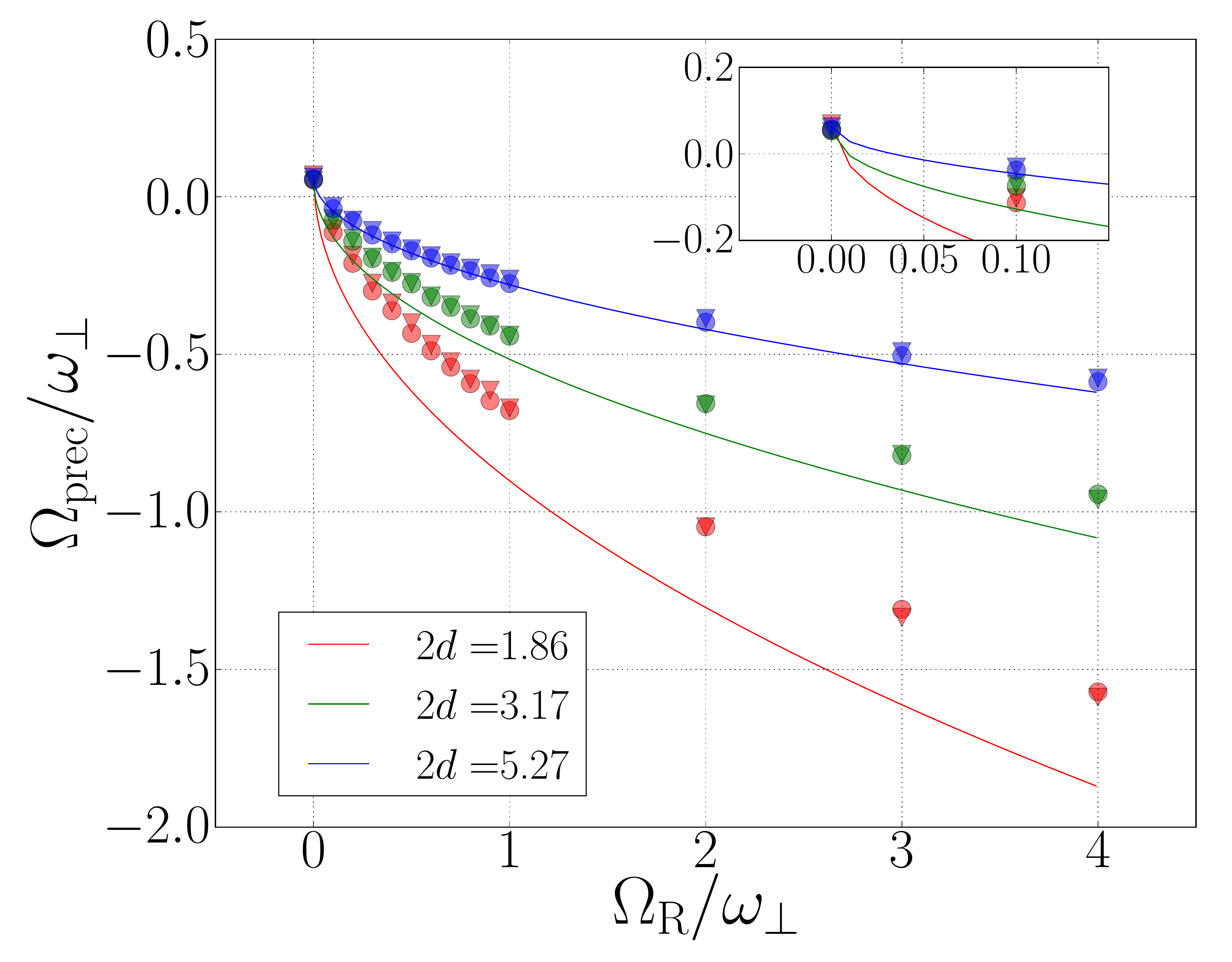}
\caption{(color online) Dependence of $\Omega _{\mathrm{prec}}$ on the Rabi coupling $\Omega_{\mathrm{R}}$ for different values of the vortex separation $2d$. The numerical solution of the GP equations (component 1 - bullets, component 2 - triangles) are in a good agreement with the analytical expression, Eq.~(\protect\ref{eq.prec}), (solid lines) for large values of $2d$ (in units of $\xi_{ho}$). The inset is a zoom of the figure for small values $\Omega_{\mathrm{R}}$, where $\Omega_{\mathrm{prec}}$ changes sign.} 
\label{fig3}
\end{figure}

\begin{figure}
\centering
\includegraphics[width=.98\columnwidth, clip=true]{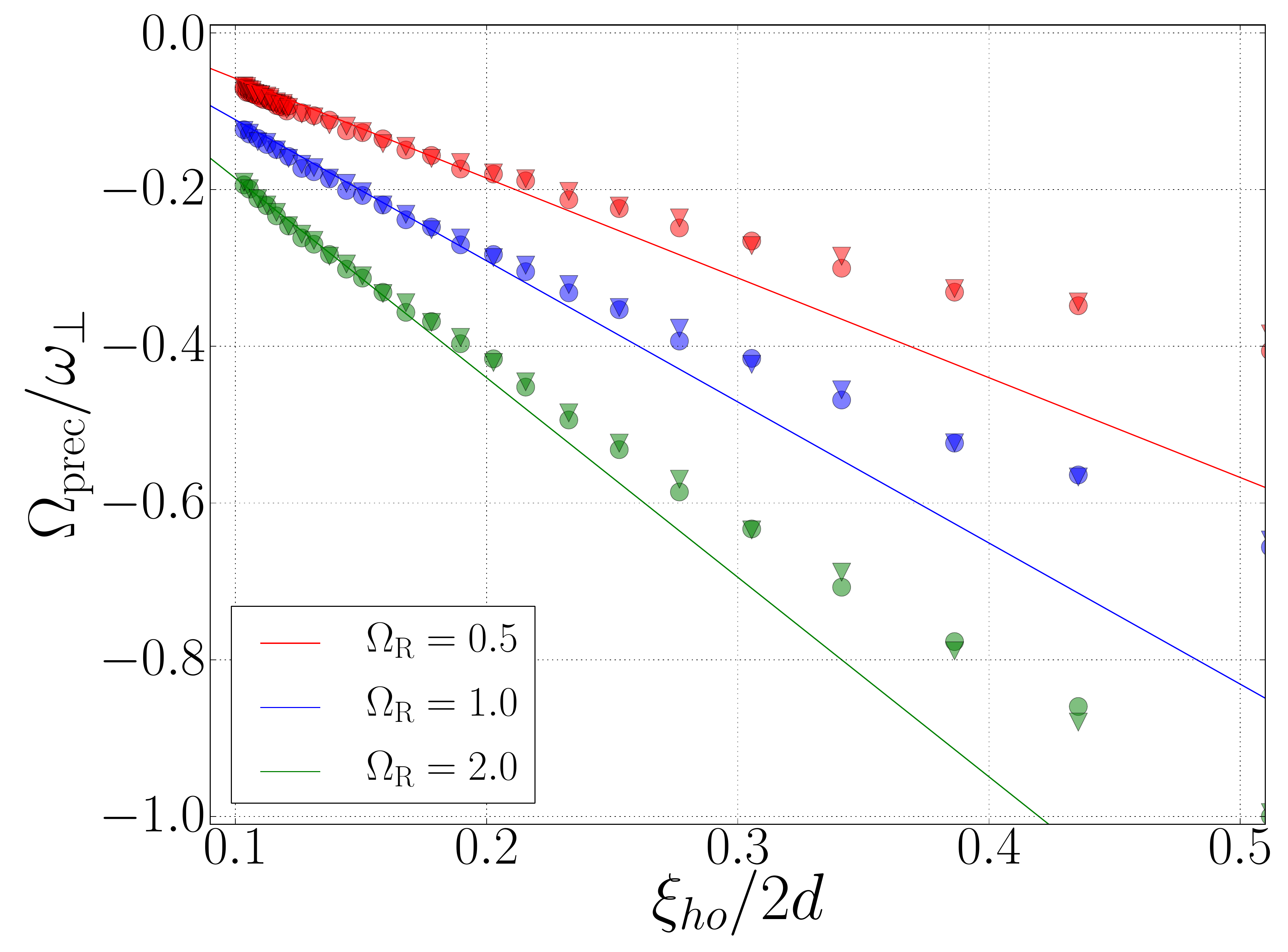}
\caption{(color online) Dependence of the  precession frequency $\Omega_{\mathrm{prec}}$ (component 1 - bullets, component 2 - triangles) on $1/2d$ (in units of $\sqrt{m\protect\omega _{\perp }/\hbar }$). The solid lines correspond to the prediction of Eq.~(\ref{eq.prec}). As in the Fig.~\ref{fig3}, the agreement is good as long as the distance $2d$ separating the two vortices is sufficiently large. }
\label{fig4}
\end{figure}

By solving the GP equation in real time one can investigate
the precession of the vortex pair and evaluate the precession
frequency $\Omega _{\mathrm{prec}}$, whose dependence on the Rabi coupling $\Omega_{\mathrm{R}}$ is
reported in Fig.~\ref{fig3} for different values of the vortex size $d$. 
Its dependence on $d$, for a fixed value of $\Omega_{\mathrm{R}}$, is instead shown in Fig.~\ref{fig4}. The results of the numerical calculations are found to agree reasonably well with the predictions of the macroscopic model~(\ref{eq.prec}) discussed in the first part of the paper (see solid lines in the figures). As expected, the discrepancies become smaller if the width of the domain wall becomes much smaller compared to the vortex separation $d_{wall} \ll 2d$. 

\begin{figure}
\centering
\includegraphics[width=.8\columnwidth, clip=true]{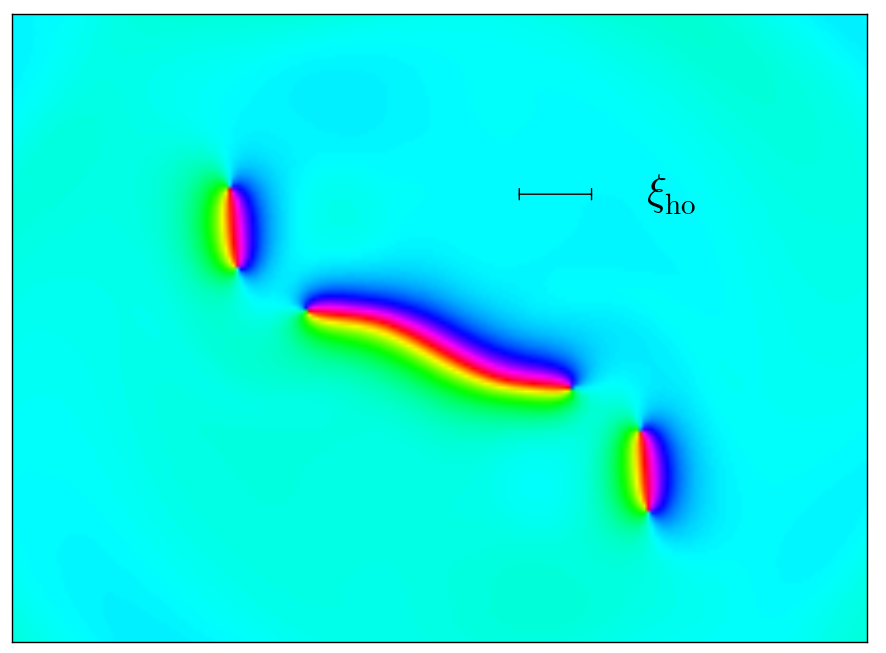}
\caption{(color online) Fragmentation of the domain wall after the evolution time $\omega_\perp t = 1.5$. One can see fragments of different size, connecting vortices of different components. The GP simulation was carried out choosing $g_{12} = 0$ and $\Omega_{\mathrm{R}} = 6 \omega_\perp$. The initial condition corresponded to a domain wall between the vortex pair of length $2d \approx 5 \xi_{\rm ho}$, symmetric with respect to the trap's center. The figure is a zoom of the central region of the cloud. }
\label{fig6}
\end{figure}

For large values of $\Omega_{\rm R}$ a long domain wall may decay into smaller fragments, resulting in the creation of new vortex pairs at the ends of these new fragments. This fragmentation is an analogue of string breaking in quantum chromodynamics. The probability of such a fragmentation becomes larger and larger as $\Omega_{\rm R}$ grows. In Fig.~\ref{fig6} we show the result of the fragmentation of a domain wall obtained at $\Omega_{\mathrm{R}} = 6 \omega_\perp$ and $g_{12} = 0$. The initial configuration -- a domain wall symmetric with respect to the center of the trap -- was then allowed to evolve through the time $\omega_\perp t = 1.5$. The figure clearly shows three fragments of various sizes, connecting vortices of different components.

\begin{figure}
\centering
\includegraphics[width=.75\columnwidth, clip=true]{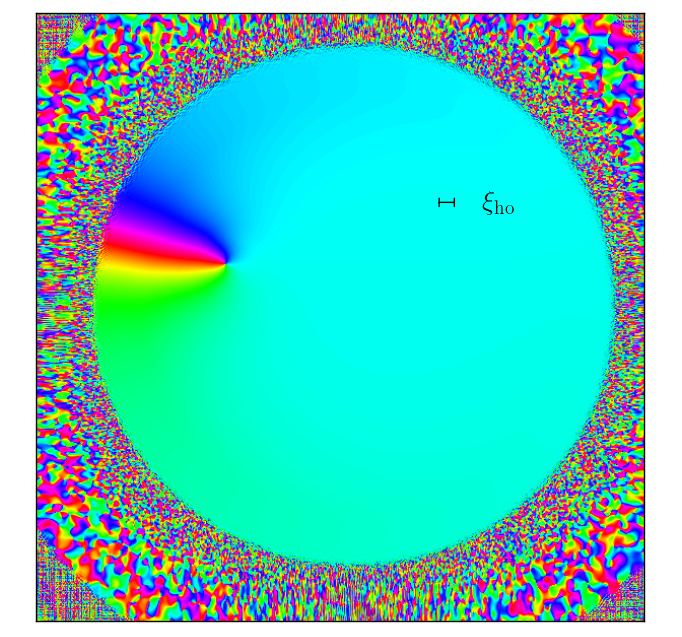}
\caption{(color online) Relative phase distribution around a single half-vortex in a two component coherently-coupled system. The vortex builds a domain wall that is attached to the nearest point of the edge of the cloud. The  phase distribution corresponds to the evolution time of $\omega_\perp t = 0.2$.  Then, the vortex starts precessing and induces the appearance of a second vortex in the component 2.}  
\label{fig5}
\end{figure}

In the introduction we emphasized the fact that, in a uniform, infinite system in the presence of the Rabi coupling, vortices cannot exist alone as single objects but only in pairs. In a finite system, such as in the presence of a confinement, single vortex lines can also exist, as the domain wall will cost a finite amount of energy, fixed by the size of the atomic cloud. We have explored single vortex configurations by considering a vortex line (corresponding to the component 1) located at some distance from the center of the trap. The domain wall which minimizes the energy corresponds to
the shortest line connecting the vortex to the external region outside the Thomas-Fermi radius, where the density of the atomic cloud is vanishing (see Fig.~\ref{fig5}). When solving the GP equations~(\ref{GP}) in real time, the vortex line in component 1 exhibits the precession according to the macroscopic prediction (\ref{Ed})-(\ref{precession}) with $2E_{v}$ replaced by $E_{v}$ and $E_{wall}$ calculated along the domain wall. However, we soon find the appearance of a second vortex in component 2, attached to the second end of the wall and emerging from the border, where its energy cost is vanishingly small. The two vortices then start rotating around each other. Eventually the original vortex of the component 1 reaches the border of the atomic cloud to disappear and reappear again after a while (an analogous behavior of vortices in a toroidal trap was observed in Ref.~\cite{Gallemi2015a}).

The results presented above were obtained assuming the inter-component interaction parameter to be $g_{12} = 0$. If $g_{12}$ is small compared to the  intra-component coupling constants, $g_{11} = g_{22} \equiv g$, we find that the influence of $g_{12}$  on the precession is almost negligible, the main role being played by the  long-range surface tension force. This behavior is consistent with the fact that stable molecules have a very small size, even for relatively large values of $g_{12}$. If $g_{12}$ is close to $g$, one can identify a critical value for the Rabi coupling, given by the expression $\Omega_{\mathrm{crit}} = \frac{1}{3} (g - g_{12}) n / \hbar$ \cite{Son2002}, above which the domain wall becomes unstable and the solution of the GP equations corresponds to a local maximum of energy, rather than to a local minimum. 


\section{Conclusions}
We expect that our predictions for the precession of half-vortex pairs and for the fragmentation of the corresponding domain wall at large Rabi coupling will stimulate new measurements on coherently coupled BECs. Experimentally, pairs of  half-vortices, connected by a domain wall, can be created by the proper imprinting of the relative phase of the two condensates. The shape of the domain wall connecting the two vortical lines is in principle observable using heterodyne methods giving rise to visible interference in the domain wall region. The precession effect could be measured using real-time detection techniques (see, for example,~\cite{Freilich2010}).

\begin{acknowledgments}
We are grateful to Franco Dalfovo, Alexander Fetter, Gabriele Ferrari and Chunlei Qu for stimulating discussions. This work was supported by ERC through the QGBE grant,  by the  QUIC grant of the Horizon2020 FET program and  by Provincia Autonoma di Trento. A.R. acknowledges support from the Alexander von Humboldt foundation. This work was supported in part by the PL-Grid infrastructure (M.T.). 
\end{acknowledgments}

\end{document}